\documentclass[aps,prl,reprint,superscriptaddress,footinbib]{revtex4-2}
\usepackage[normalem]{ulem}
\usepackage[T1]{fontenc}
\usepackage{amsmath,amssymb,mathtools}
\usepackage{xcolor}
\usepackage{booktabs}
\usepackage{graphicx}
\usepackage{microtype}
\usepackage[hidelinks]{hyperref}

\usepackage{braket}

\newcommand{\F}{\mathbb{F}}
\newcommand{\Tgate}{\mathrm{T}}
\newcommand{\CS}{\mathrm{CS}}
\newcommand{\CCZ}{\mathrm{CCZ}}
\newcommand{\RM}{\operatorname{RM}}
\newcommand{\wt}{\operatorname{wt}}

\usepackage{verbatim}

\usepackage{array,longtable}

\makeatletter
\appdef\@mkpream@relax{\let\@endpbox\relax}
\makeatother
\newcommand{\FieldOutputFive}{\mathrm{CS}_{45}\allowbreak\mathrm{CCZ}_{135}\allowbreak\mathrm{CCZ}_{234}}
\newcommand{\FieldOutputSix}{\mathrm{CS}_{12}\allowbreak\mathrm{CCZ}_{134}\allowbreak\mathrm{CCZ}_{135}\allowbreak\mathrm{CCZ}_{146}\allowbreak\mathrm{CCZ}_{234}\allowbreak\mathrm{CCZ}_{256}}

\begin{document}

\title{Classification of Generalised Triorthogonal Codes through Length 54}

\author{Adam Wills}
\email{a_wills@mit.edu}
\affiliation{Center for Theoretical Physics --- a Leinweber Institute, Massachusetts Institute of Technology, Cambridge, MA}
\affiliation{IBM Research, IBM T. J. Watson Research Center, NY}
\author{Shubham P. Jain}
\email{shubhamj810@gmail.com}
\affiliation{IBM Research, IBM T. J. Watson Research Center, NY}
\affiliation{Joint Center for Quantum Information and Computer Science, NIST, University of Maryland, College Park, Maryland 20742, USA}
\author{Shraddha Singh}
\email{shraddha.s@ibm.com}
\affiliation{IBM Research, IBM T. J. Watson Research Center, NY}
\date{\today}

\begin{abstract}
Magic state distillation is a widely-considered primitive in fault-tolerant quantum computation for the preparation of high-fidelity non-Clifford resources. The most commonly-considered class of such protocols are generalised triorthogonal codes; these distil $n$ noisy input $\mathrm{T}$ states into purified third-level diagonal magic states. Extensive prior work has searched this space of protocols, often using heuristic methods that do not guarantee optimality. Existing classification work is limited to protocols distilling $k$ outputted $\mathrm{T}$ states with $n+k\leq 38$~\cite{nezami2022classification}. In this work, we significantly expand this classification to all protocols with lengths $n \leq 54$. We restrict to distance $d \geq 3$ to keep the classification of a sensible size, and since efficient searches are well-understood at distance $2$~\cite{Singh2026BorrowedIdentities}. Moreover, our results are complementary to synthillation~\cite{campbell2017unified}, which can distil third-level states from $\mathrm{T}$ states, but only at distance $2$.
Under optimality in terms of input count, space footprint, and distance for a given output, we find $74$ optimal generalised triorthogonal protocols in our range, $65$ of which are new to the literature.
To achieve our classification, we extend the classification of unital triorthogonal spaces of Nezami and Haah~\cite{nezami2022classification} from length $38$ to $54$ using a directional derivative method. Using these as the stabiliser spaces, we add logical rows that satisfy the triorthogonality constraints to create full protocols.
\end{abstract}

\maketitle
Stabiliser operations are frequently viewed as the cheapest operations to execute in a universal fault-tolerant architecture, since it is often most natural to construct quantum codes supporting them transversally, or via code surgery techniques. While they are not universal, they become universal when supplemented with a magic state
\begin{equation}
    \lvert U\rangle=U\lvert +\rangle^{\otimes k},
\end{equation}
where $U$ is a $k$-qubit diagonal non-Clifford gate~\cite{bravyi2005universal}. As long as $U$ is in the third level of the Clifford hierarchy,~\cite{gottesman1999demonstrating}, $\ket{U}$ may be consumed to teleport $U$ to another $k$-qubit quantum state using only stabiliser operations. Throughout, we exclusively consider diagonal, third-level gates; examples include $\mathrm{T}$, $\mathrm{CS}$, $\mathrm{CCZ}$, and their products. It turns out that these three gates generate all third-level diagonal gates~\cite{cui2017diagonal}.

Magic state distillation (MSD) is a leading method for implementing non-Clifford operations in fault-tolerant quantum computing architectures. Using Clifford operations alone, many copies of noisy magic states may be \emph{distilled} to a smaller number of higher-fidelity magic states, possibly of a different type~\cite{bravyi2005universal,bravyi2012magic}. The production of high-fidelity non-Clifford resources has long been identified as a key bottleneck in fault-tolerant quantum computing architectures. It is therefore well-motivated to construct high-performance MSD protocols. We want protocols with
\begin{enumerate}
    \item High distance. This is the minimum number of input states that must be errorful to create some undetectable logical error on the output; this leads to good error suppression.
    \item For a fixed input, the scheme should produce the most useful non-Clifford resource as possible. For example, restricting to schemes distilling $n$ $\ket{\mathrm{T}}$ states to $k$ $\ket{\mathrm{T}}$ states, we simply desire the greatest ratio $\frac{k}{n}$.
    \item The scheme has a small spatial footprint; this is the number of qubits on which it may be run. Note that these qubits are the logical qubits of some host code supporting fault-tolerant stabiliser operations.
\end{enumerate}
Complementary magic preparation methods such as magic state cultivation~\cite{gidney2024magic,hirano2025efficient,gupta2024encoding,rosenfeld2025magic,jacoby2025magic,itogawa2025efficient,sahay2026fold} are very efficient in that they produce non-Clifford resources with physical-level qubits, rather than logical-level. However, these methods typically require a large physical spacetime volume to produce one $\ket{\mathrm{T}}$ state, and do not get outputs at extremely low error rates. It is therefore well-motivated to consider MSD, either on directly injected magic states, or on states that have been cultivated by a small amount.

The most commonly-considered framework for MSD is that of generalised triorthogonal codes~\cite{haah2018codes}; relatively few protocols are known to lie outside this class. These codes arise from generalised triorthogonal matrices, defined below, which define a quantum CSS code with the appropriate structure for distilling $n$ noisy input $\ket{\mathrm{T}}$ states to another diagonal gate in the third level of the Clifford hierarchy. These protocols are restricted in the sense that they only take $\ket{\mathrm{T}}$ states as input. On the other hand, the injection of $\ket{\mathrm{T}}$ states into surface codes and qLDPC codes, which could be the host code for the distillation scheme, is more widely studied than for other states, and appears more simple than for multi-qubit gates. 

In this work, we classify generalised triorthogonal codes to length $n \leq 54$, at distance $d \geq 3$. Our results go beyond the existing classification due to Nezami and Haah~\cite{nezami2022classification} of triorthogonal codes~\cite{bravyi2012magic} producing $k$ $\ket{\mathrm{T}}$ state outputs, up to length $n \leq 38-k$. Our restriction to distance $d \geq 3$ is made because the protocols are well-understood at distance $2$~\cite{Singh2026BorrowedIdentities}, and because there would be an impractical number of protocols if we were to allow distance $2$ protocols in the classification. We are motivated to consider generalised triorthogonality, rather than just triorthogonal codes~\cite{bravyi2012magic}, which distil $n$ $\ket{\mathrm{T}}$ states to $k$ $\ket{\mathrm{T}}$ states, as we would particularly like to consider the distillation of exotic magic states. It is often more efficient to distil a specialised state for the given algorithmic task than to distil simple states like $\mathrm{T}$, and synthesise the operation~\cite{campbell2016efficient,campbell2017unified}. Notably, synthillation~\cite{campbell2017unified} can systematically construct generalised triorthogonal codes for a particular magic state, but is restricted to distance $2$; in this sense, our results are complementary to synthillation.
\begin{figure*}[t]
    \centering
    \includegraphics[width=0.98\textwidth]{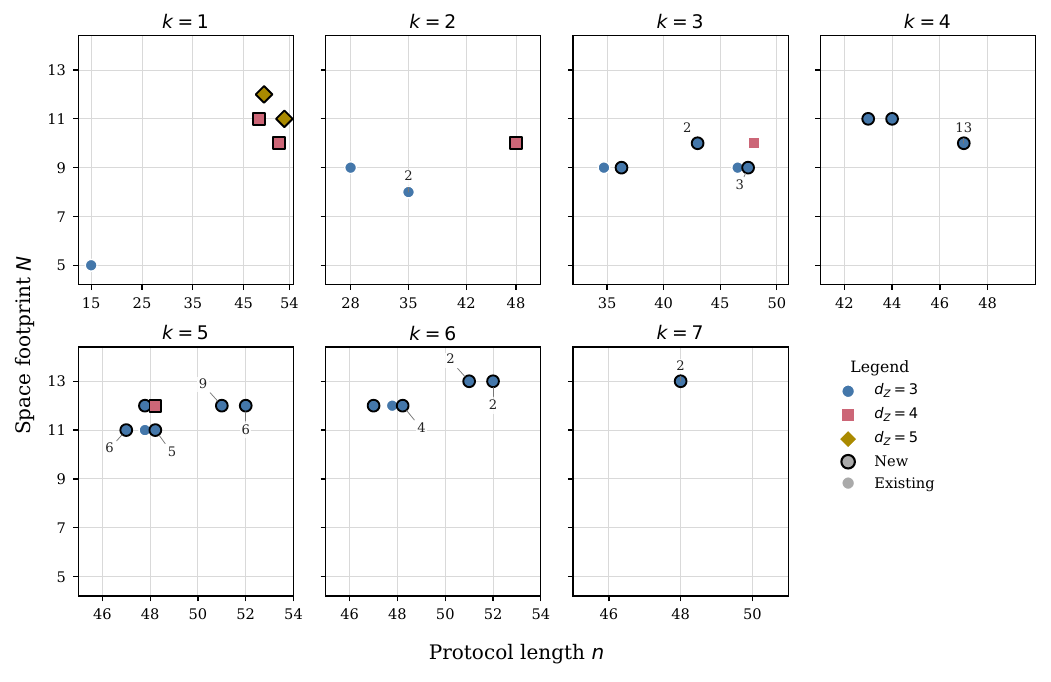}
    \caption{Complete classification of generalised triorthogonal protocols with $n\leq54$ and $d_Z\geq3$, under $\mathrm{CNOT}+\mathrm{S}$ output equivalence, and discarding those that are worse than another protocol with the same output in their distance, input $\ket{\mathrm{T}}$ count, and space footprint $N$. Each panel fixes the output qubit count $k$. We do not state the outputted $k$-qubit magic state class here; full details are given in Table~\ref{tab:complete-pareto}. In general, the outputted magic state is a different class to $\ket{\mathrm{T}}^{\otimes k}$. Solid black outlines mark new points; unoutlined markers denote existing points, including those obtained by simple operations on existing protocols. Numerical labels count multiple protocols with inequivalent outputs that are otherwise the same; an unlabelled marker represents one protocol. Markers that would otherwise be coincident are slightly separated for visibility. We emphasise that the definition of space footprint $N$ here is the number of rows in the generalised triorthogonal matrix. For certain protocols, a lower space footprint is possible using protocol compression/qubit recycling~\cite{xu2026distillingmagicstatesbicycle,jacinto2026compact}, which use additional mid-circuit measurements. As one concrete example,~\cite{jacinto2026compact} shows that Bravyi and Haah's $49$-to-$1$ $\ket{\mathrm{T}}$ state protocol at distance $5$ \cite{bravyi2012magic} may be executed with a space footprint of $5$ qubits.}
    \label{fig:frontier}
\end{figure*}

\emph{Generalised triorthogonal matrices.---}
Let $\F_2=\{0,1\}$ be the binary field, and consider a matrix
\begin{equation}
    G=\begin{bmatrix}G_1\\ \hline G_0\end{bmatrix}\in\F_2^{N\times n}.
    \label{eq:block-matrix}
\end{equation}
The $k$ rows of $G_1$ are called \emph{logical rows}; the $r=N-k$ rows of $G_0$ are called \emph{stabiliser rows}. For $a \in [N]$, the $a$'th row of $G$ is denoted $g^a$, and the overlap of a collection of three rows $g^{a_1}, g^{a_2}, g^{a_3}$ is
\begin{equation}
    \langle g^{a_1}, g^{a_2}, g^{a_3}\rangle
      =\sum_{j=1}^{n}\prod_{i=1}^{3}G_{a_ij}\pmod 2,
    \label{eq:overlap}
\end{equation}
for $a_i \in [N]$ not necessarily distinct. We call $G$ generalised triorthogonal when any such overlap is zero, if the overlap involves a stabiliser row (if some $a_i > k$). The remaining logical overlaps yield a symmetric trilinear form $\tau: \mathbb{F}_2^k \times \mathbb{F}_2^k \times \mathbb{F}_2^k \to \mathbb{F}_2$,
\begin{equation}
    \tau(u,v,w)=\sum_{j=1}^{n}(uG_1)_j(vG_1)_j(wG_1)_j.
    \label{eq:tensor}
\end{equation}
A generalised triorthogonal $G$ may be used to define an MSD protocol taking $n$ $\ket{\mathrm{T}}$ states as input, and outputting the third-level diagonal gate
\begin{equation}
 U_\tau=\prod_a\Tgate_a^{\tau_{aaa}}
       \prod_{a<b}\CS_{ab}^{\tau_{aab}}
       \prod_{a<b<c}\CCZ_{abc}^{\tau_{abc}}.
 \label{eq:gate}
\end{equation}
In this paper, two third-level diagonal magic states are considered equivalent if they are equal up to the action of the Clifford gates $\mathrm{CNOT}$ and $\mathrm{S}$. This is a reasonably computationally tractable equivalence via the connection to phase polynomials~\cite{amy2019tcount} (although still challenging for larger magic states). We may talk about the \textit{class} of a (third-level, diagonal) magic state as its equivalence class under this equivalence.

We define the \emph{length} of the protocol corresponding to $G$ as $n$, the number of columns of the matrix, which is also the number of inputted $\mathrm{T}$ states. Its \emph{space footprint} is the total row count $N=k+r$, since the protocol may always be run on this many (logical) qubits~\cite{GOSC}. Since $\mathrm{X}$ errors may be twirled~\cite{bravyi2012magic}, the distance relevant to magic state distillation is
\begin{equation}
 d = d_Z=\min\{\wt(v):G_0v=0,\;G_1v\neq0\},
 \label{eq:distance}
\end{equation}
where $v\in\F_2^n$ and $\wt(v)$ is its number of nonzero entries. Thus $d_Z$ is the minimum number of input errors that can pass all stabiliser checks while changing the logical output. Consider a protocol (corresponding to a generalised triorthogonal matrix) that distils a given output class, with fixed $d_Z$, suppose it has length and space footprint $(n,N)$. We call the protocol \emph{Pareto optimal} or \emph{Pareto dominant} if no other protocol is at least as good in both coordinates $(n,N)$ and strictly better in one. We will classify this Pareto frontier up to lengths $n \leq 54$, with distances $d_Z \geq 3$. We further restrict to matrices with pairwise distinct columns and no all-zeros column, since a pair of equal columns, or a zero column, could be excluded to produce a better protocol.

\emph{Reduction to unital triorthogonal spaces.---}A unital triorthogonal space of length $c$ is a linear subspace of $\mathbb{F}_2^c$ containing the all-ones vector (unitality), and in which the overlap of any three (not necessarily distinct) vectors, as in Equation~\eqref{eq:overlap}, is zero (triorthogonality). Notice that a unital triorthogonal space must have even length. At a high level, our strategy is to reduce the classification at hand to a classification of unital triorthogonal spaces.

Let us show how we pass from a generalised triorthogonal matrix to a unital triorthogonal space, and back. Given a generalised triorthogonal matrix $G$, write column $j$ as $(\lambda^{(j)},x^{(j)})\in\F_2^k\times\F_2^r$, and forget the logical rows given by $\lambda^{(j)}$. Our distance requirement forces $x^{(i)} \neq x^{(j)}$ for $i \neq j$ and $x^{(i)} \neq 0$ for all $i$. 
\begin{equation}
    X\coloneq\{x^{(1)},\ldots,x^{(n)}\}\subseteq\F_2^r
    \label{eq:support}
\end{equation}
is thus a set, rather than a multiset. When $n$ is odd, we add the zero vector to $X$ to produce a new set $Y$ of even cardinality; when $n$ is even, let $Y=X$. The matrix with columns 
\begin{equation}\label{eq:unital_triortho_space_cols}
    \{(1,y)^T:y \in Y\},
\end{equation}
generates (meaning its rowspan is) a unital triorthogonal space, which follows from the stabiliser overlap conditions. 

On the other hand, given a unital triorthogonal space of dimension $1+r$, we write down a generator matrix, where we always write its first row as the all-ones vector. We must consider all the ways to recover a stabiliser matrix from the generator matrix for the unital triorthogonal space, that is, all the ways to reverse the construction above. First, we may simply take the generator matrix for the unital triorthogonal space to be the stabiliser matrix. This corresponds to cases above where adding the all-ones row did not increase ths size of the space above. On the other hand, we need also consider cases where the all-ones row was a legitimate addition. Here, we must consider every $o \in \mathbb{F}_2^r$, and add the all-ones row to the lower $r$ rows corresponding to the support of $o$. Remove the all-ones row. If some all-zero column remains, remove that too, and what is left is a candidate stabiliser matrix. Adding the all-ones row into all the other rows in every possible way effectively reverses the row operations that could have taken place on the generator matrix for the unital triorthogonal space, while preserving the top all-ones row.

Given a $G_0$, to get all possible $G$, we want to consider all the sets of $k$ logical rows that can be added such that the result is a generalised triorthogonal matrix. The notion of constructing protocols by adding compatible logical rows to a valid stabiliser space has appeared in the literature in certain forms before~\cite{haah2018codes,gong2026extension}, but we approach this idea more systematically as follows. The overlap conditions (see Equation~\eqref{eq:overlap}) involving exactly one logical row define a linear space of candidate rows; we quotient this space by the stabiliser space $\text{rowspan}(G_0)$, since logical rows must be identified if they differ by a sum of stabilisers. Working in the quotiented space, it remains to enforce the overlap conditions involving two logical rows. We convert this into a condition on bilinear forms, where each bilinear form is defined by the overlap with a single stabiliser row. Allowed sets of $k$ logical rows correspond to $k$-dimensional subspaces that are \emph{totally isotropic} with respect to these bilinear forms, meaning these bilinear forms vanish identically on the subspace. See the Supplemental Material for details. 

Before moving on, we comment that the puncturing-based reduction from triorthogonal matrices to unital triorthogonal spaces due to Nezami and Haah~\cite{nezami2022classification} is distinct to ours, and in particular leads to a shorter protocol than the unital triorthogonal space. In particular, by classifying unital triorthogonal spaces up to length $38$, they are able to classify triorthogonal protocols with $n$ inputs and $k$ $\ket{\mathrm{T}}$-state outputs up to $n+k\leq 38$. We will classify unital triorthogonal spaces up to length $54$, and this will give us a classification of generalised triorthogonal matrices with $n \leq 54$.

\emph{Unital Triorthogonal Space Classification using Directional Derivatives.---}
To classify unital triorthogonal spaces up to length $38$, Nezami and Haah use a connection to Reed-Muller codewords, which we also use. Every function $f:\mathbb{F}_2^m \to \mathbb{F}_2$ has a unique expression as a polynomial over $\mathbb{F}_2$ in the variables $\{x_i:i \in [m]\}$, where $x_i^2 = x_i$, and its degree is the maximum number of variables in a monomial in that expression. An equivalence relation on such functions is established by considering affine relations
\begin{equation}
    \vec{x} \mapsto A\vec{x}+b,
\end{equation}
where $\vec{x} = (x_i)_{i=1}^m$ is a vector of variables, $A$ is an $m \times m$ invertible binary matrix, and $b \in \mathbb{F}_2^m$, and we note that affine equivalent functions have the same weight, where their weight is the size of their support over $\mathbb{F}_2^m$. Nezami and Haah show that unital triorthogonal spaces of dimension $1+m$, with length $c$, are in one-to-one correspondence with affine equivalence classes of such functions of degree $\leq m-4$ and weight $c$. To ensure that the unital triorthogonal space has dimension $1+m$ and not lower, one ensures that the function is not divisible by a nonconstant function of degree one; see the Supplemental Material for details. 

Classifying unital triorthogonal spaces thus becomes a problem of classifying Reed-Muller codewords in $\RM(m-4,m)$ up to affine equivalence, of each weight (the affine equivalence corresponds to the fact that the unital triorthogonal space is invariant under row operations on its generator matrix that preserve the top all-ones row). Nezami and Haah are able to do this up to length (weight) $38$ \footnote{We comment that two unital triorthogonal spaces are missing from their catalogue; see the accompanying data~\cite{classificationdata}.} using classical results on Reed-Muller codewords~\cite{kasami1970weight,kasami1976weight}.

In this work, we develop a \emph{directional derivative} method for Reed-Muller codeword classification, thus pushing the classification to length $54$. This is distinct from, but inspired by, results establishing asymptotic bounds on Reed-Muller weight distributions using directional derivatives~\cite{kaufman2012weight,abbe2015reed,sberlo2020performance}. Let $f$ be a polynomial as above and let $z \coloneq (x_1, \ldots, x_{m-1})$; we may uniquely write
\begin{equation}
    f(z, x_m) = f_0(z) + x_m\cdot g(z),
\end{equation}
where 
\begin{align}
    f_0(z) &= f(z, 0)\\
    f_1(z) &=  f(z, 1)\\
    g&= f_0 + f_1,
\end{align}
and note $\deg(g) \leq m-5$. The idea is that $g$ has a smaller number of variables, and a lower degree, and so we may aim to classify codewords recursively; we must however worry about the weight of this polynomial.

Let
\begin{align}
    P &\coloneq \{z \in \mathbb{F}_2^{m-1}: f(z,0)\ne f(z,1)\},\\
    D &\coloneq \{z \in \mathbb{F}_2^{m-1}: f(z,0)=f(z,1)=1\}.
\end{align}
Notice that $c = \wt(f) = |P| + 2|D|$. We are free to perform affine transformations to redefine the variables $(z, x_m)$. For computational tractability, we imagine a transformation that maximises $|P|$ and minimises $|D|$ \footnote{$g$ is a derivative of $f$, and we consider choosing the directional derivative of $f$ with maximal weight. One intuition behind our method is that this choice enables us to reconstruct polynomials from their derivatives, and our choice of direction enables the minimal information loss from the polynomial to the derivative.}. In the Supplemental Material, we show an upper bound on what $|D|$ can be taken to be after such a transformation. We further have $\wt(g) = c-2|D|$, and we may take the codeword $g$ from some lower-variable catalogue.

Given $g$, we reconstruct all possible $f$. We consider all possible choices for $D$, which are sets of $|D|$ points in $\mathbb{F}_2^{m-1}\setminus P$. We must finally choose whether $f$ will be supported on $(z,0)$ or $(z,1)$, for each $z \in P$; in the Supplemental Material, we show that the requirement $\deg(f_0) \leq m-4$ reduces these choices to a linear system which we may solve.

\emph{Results.---}
The complete Pareto frontier, shown in Fig.~\ref{fig:frontier}, contains 74 points spanning 62 output magic classes, with logical dimension at most seven. $65$ of the $74$ protocols appear new to the literature, even when allowing basic modifications to the existing protocols (see the Supplemental Material). Of the protocols, 67 have $d_Z=3$, five have $d_Z=4$, and two have $d_Z=5$.

\emph{Outlook.---}
The accompanying data~\cite{classificationdata} contain the complete classifications of protocols and unital triorthogonal spaces, and verification evidence. All theory and code necessary for regeneration is supplied in the repository~\cite{repo}. The present classification required of the order of $10^5$ core-hours of computation. A rough estimate of a length-56 classification with the same method is $10^8$ core-hours~\cite{repo}, suggesting the need for additional theoretical results to push the classification further, rather than just more computation. To characterise this blowup, we note that the numbers of unital triorthogonal spaces at lengths $48, 50, 52$, and $54$ are $28,875$, $304,611$, $7,519,688$, and $293,172,583$, respectively \cite{classificationdata}. This suggests that further protocol classification should not proceed via unital triorthogonal space classification. To this end, non-exhaustive, yet relatively fast methods for discovering magic state distillation protocols, may be found in our companion paper~\cite{jain2026symmetry}.

Aside from further classification, a greater understanding of the magic states we can produce, and their utility for particular algorithmic tasks, would be very valuable. As one concrete example, a better understanding of the equivalence of magic states under full stabiliser equivalence, rather than simply our notion of $\mathrm{CNOT}+\mathrm{S}$-equivalence, is needed; certain states are known to be equivalent under full stabiliser operations, but not $\mathrm{CNOT}$ and $\mathrm{S}$, such as the magic states for $\mathrm{CS}_{12}\mathrm{CS}_{13}$ and $\mathrm{CCZ}_{123}$~\cite{howard2017application}. As another example,~\cite{jain2026symmetry} presents a protocol with $43$ inputs, distance $3$, space footprint $11$, and a $k=4$-qubit output which is Clifford equivalent, but not $\mathrm{CNOT}+\mathrm{S}$-equivalent, to $\mathrm{T}_1\mathrm{CCZ}_{234}$. Referring to our catalogue in Table~\ref{tab:complete-pareto}, we see that to obtain the same output at the same distance, we require $n = 47$ inputs, albeit with space footprint $N = 10$.

\emph{Acknowledgements.---} The authors thank Anqi Gong for inspiring discussions. A.W. acknowledges support from the MIT-IBM Watson AI Lab. This preprint is assigned number MIT-CTP/6111.
The key conceptual ideas behind the classification were developed by the authors. The authors acknowledge the use of ChatGPT 5.6 Sol and ChatGPT 6 Astra for assistance in implementing theoretical ideas in code~\cite{repo}, and making the figure. The authors acknowledge the use of the same AI tools for the development of further conceptual ideas that reduce the search space for the classification~\cite{repo}. 
This research used resources of the National Energy Research Scientific Computing Center, a DOE Office of Science User Facility supported by the Office of Science of the U.S. Department of Energy under Contract No. DE-AC02-05CH11231 using NERSC award NERSC DDR-ERCAP0038585. We acknowledge the MIT Office of Research Computing and Data for providing high-performance computing resources that have contributed to the research results reported within this paper. The computations in this paper were partially run on the FASRC Cannon cluster supported by the FAS Division of Science Research Computing Group at Harvard University. This work is supported by the National Science Foundation under Cooperative Agreement PHY-2019786 (The NSF AI Institute for Artificial Intelligence and Fundamental Interactions, http://iaifi.org/). This research was done in part using services provided by the OSG Consortium~\cite{osg_ospool_2006,ruth2007open,sfiligoi2009pilot,osg_osdf_2015}, which is supported by the National Science Foundation awards \#2030508 and \#2323298.

\bibliography{references}

\clearpage
\onecolumngrid
\setcounter{equation}{0}
\renewcommand{\theequation}{S\arabic{equation}}
\renewcommand{\theHequation}{supp.\arabic{equation}}
\setcounter{table}{0}
\renewcommand{\theHtable}{supp.\arabic{table}}
\setcounter{section}{0}
\section*{Supplemental Material}

In this Supplemental Material, we give further details of the classification of generalised triorthogonal codes through length $54$, at distance $d_Z\geq3$. We use the same definitions as in the main text: $n$ is the number of inputted $\ket{\mathrm{T}}$ states, $k$ is the number of output logical qubits, and $N$ is the number of matrix rows. Output magic states are identified up to equivalence by $\mathrm{CNOT}$ and $\mathrm{S}$ gates. We first explain the reduction to unital triorthogonal spaces and the construction of compatible logical rows. We then describe the directional derivative method for classifying the spaces. Finally, we give the full Pareto frontier in Table~\ref{tab:complete-pareto}, together with its comparison to the literature. The accompanying data~\cite{classificationdata} contain the matrices, space catalogues and verification evidence; the repository~\cite{repo} contains the theory and code for the classification.

\section{Generalised triorthogonal matrices}

Let
\begin{equation}
G=\begin{bmatrix}G_1\\ \hline G_0\end{bmatrix}\in\F_2^{(k+r)\times n}
\end{equation}
have full row rank, with $N=k+r$. The first $k$ rows are logical rows and the remaining $r$ are stabiliser rows. As in the main text, we denote row $a$ by $g^a$ and define
\begin{equation}
\langle g^{a_1},g^{a_2},g^{a_3}\rangle
=\sum_{j=1}^n G_{a_1j}G_{a_2j}G_{a_3j}\pmod2.
\label{eq:supp-overlap}
\end{equation}
The row indices $a_1, a_2, a_3$ need not be distinct. We call $G$ generalised triorthogonal when this overlap vanishes whenever at least one of the rows is a stabiliser row. The remaining overlaps define a symmetric trilinear form
\begin{equation}
\tau(u,v,w)=\sum_{j=1}^n(uG_1)_j(vG_1)_j(wG_1)_j,
\qquad u,v,w\in\F_2^k.
\end{equation}
This is a symmetric trilinear form, with $\tau(u,u,v)=\tau(u,v,v)$. The triorthogonal matrix $G$ defines a magic state distillation protocol with $n$ input states and output state $\ket{U_\tau}=U_\tau\ket{+}^{\otimes k}$, where
\begin{equation}
U_\tau=\prod_a\Tgate_a^{\tau_{aaa}}
\prod_{a<b}\CS_{ab}^{\tau_{aab}}
\prod_{a<b<c}\CCZ_{abc}^{\tau_{abc}},
\end{equation}
see~\cite{haah2018codes}. We identify matrices under permutations of columns, changes of stabiliser or logical basis, and additions of stabiliser rows to logical rows. We remove dependent stabiliser rows and take logical rows that are independent modulo the stabiliser space. A logical basis change acts on $\tau$ by $\mathrm{GL}(k,2)$, and diagonal Clifford corrections do not change its class. This gives the $\mathrm{CNOT}+\mathrm{S}$ output equivalence used in the main text, not full Clifford equivalence. In particular,
\begin{equation}
\mathrm{S}_i\mathrm{S}_j\mathrm{CNOT}_{i\to j}\mathrm{S}_j^{-1}\mathrm{CNOT}_{i\to j}
=\mathrm{CZ}_{ij},
\end{equation}
so controlled-$\mathrm{Z}$ corrections are included, but Hadamard gates are not. The phase-polynomial framework of Ref.~\cite{amy2019tcount} uses this restricted gate setting.

We restrict to matrices with pairwise distinct columns and no all-zero column, as in the main text. The reason is that either two identical columns or an all-zero column could be deleted to produce a better protocol. Pairwise distinctness is also called \emph{projectivity}. The distance is
\begin{equation}
d_Z=\min\{\wt(v):G_0v=0,\;G_1v\neq0\},\qquad v\in\F_2^n.
\label{eq:supp-distance}
\end{equation}

The spatial footprint is $N=k+r$, the total number of matrix rows, including all logical rows. This is the same footprint used in the figure and Table~\ref{tab:complete-pareto}. We do not minimise over schedules that recycle qubits through additional mid-circuit measurements. Such a schedule may use fewer active qubits than $N$~\cite{xu2026distillingmagicstatesbicycle,jacinto2026compact}, but this is a different resource measure.

\section{Reduction to unital triorthogonal spaces}

Write column $j$ of $G$ as
\begin{equation}
g_j=(\lambda^{(j)},x^{(j)})\in\F_2^k\times\F_2^r.
\end{equation}
Here $\lambda^{(j)}$ and $x^{(j)}$ are column $j$ of $G_1$ and $G_0$, respectively. The superscript $(j)$ labels the column, while $x_a^{(j)}=(G_0)_{aj}$ denotes its entry in stabiliser row $a$. Let us first forget the logical rows and consider which sets of stabiliser columns can occur.

\subsection{From stabiliser rows to an even support}

Suppose $G$ is in the stated scope and $d_Z\geq3$, and let $e_j\in\F_2^n$ denote the $j$th standard basis vector. If $x^{(i)}=x^{(j)}$ for $i\neq j$, projectivity implies $\lambda^{(i)}\neq\lambda^{(j)}$. The weight-two vector $e_i+e_j$ then has zero stabiliser syndrome and nonzero logical syndrome, contradicting the distance condition. Similarly, $x^{(j)}=0$ would force $\lambda^{(j)}=0$, since otherwise $e_j$ would be a weight-one logical error. This would be an all-zero column of $G$, which we exclude. We therefore have a set of distinct nonzero stabiliser columns,
\begin{equation}
X\subseteq\F_2^r\setminus\{0\},
\label{eq:supp-X}
\end{equation}
where $|X| = n$. Each $u\in X$ is a column of $G_0$, with $u_a$ its entry in row $a$. Thus the overlap parity of stabiliser rows $a,b,c$ is
\[
\sum_{j=1}^{n}(G_0)_{aj}(G_0)_{bj}(G_0)_{cj}
=\sum_{u\in X}u_a u_b u_c,
\]
with sums taken in $\F_2$. Repeated indices give the lower-degree monomials. Thus generalised triorthogonality implies
\begin{equation}
\sum_{u\in X}p(u)=0
\quad\text{for every square-free monomial $p$ of degree $1$, $2$ or $3$.}
\label{eq:moments}
\end{equation}
Notice that every such monomial vanishes at zero. When $n$ is odd, we may therefore add zero to $X$ without changing these sums. Define
\begin{equation}
Y=\begin{cases}
X,&n\text{ even},\\
X\cup\{0\},&n\text{ odd}.
\end{cases}
\label{eq:even-support}
\end{equation}
Then $|Y|$ is even and Equation~\eqref{eq:moments} holds for $Y$. The matrix with columns $(1,y)^T$, for $y\in Y$, has even single, pair and triple overlaps, including those involving its top all-ones row. Its row span is therefore a unital triorthogonal space, for which a generator matrix has distinct columns. In conclusion, from the stabiliser rows of a generalised triorthogonal matrix, we have been able to form a generator matrix for a unital triorthogonal space.

\subsection{Recovering stabiliser rows}

We now reverse this construction. Start with a unital triorthogonal space $\mathcal U$ (for example from our classification), of dimension $m+1$ and length $c$, with distinct generator columns. Choose a full-rank generator matrix
\[
H=\begin{bmatrix}\mathbf1\\R\end{bmatrix}\in\F_2^{(m+1)\times c},
\qquad R\in\F_2^{m\times c},
\]
where $\mathbf1$ is the all-ones row. There are two ways to construct possible stabiliser matrices $G_0$, and we must consider both when classifying generalised triorthogonal matrices.

\emph{Retain all rows.} We may take the entire generator matrix as the stabiliser matrix:
\begin{equation}
G_0=H,\qquad r=m+1,\quad n=c.
\label{eq:supp-retain-unital}
\end{equation}
Its columns are distinct and nonzero, and its row overlaps vanish because $\mathcal U$ is triorthogonal.

\emph{Remove the all-ones row.} For each column vector $b\in\F_2^m$, form
\[
R_b=R+b\mathbf1,
\]
where $b\mathbf1$ is the matrix with all columns equal to $b$, of the appropriate dimensions. Thus row $a$ of $R_b$ is row $a$ of $R$ plus $b_a\mathbf1$. If $R_b$ has a zero column, delete that column to obtain $G_0$; otherwise, take $G_0=R_b$. The columns of $R_b$ are distinct: equality of two would imply equality of the corresponding columns of $H$. Hence there is at most one zero column. The rows of $R_b$ are independent, since the rows of $H$ are independent, and deleting a zero column does not change their rank or overlaps. This construction therefore gives $r=m$, with $n=c-1$ when a column is deleted and $n=c$ otherwise. All stabiliser overlaps vanish because the rows of $R_b$ belong to $\mathcal U$.

To see that these two constructions suffice, start with any stabiliser matrix $G_0$ in the stated scope. If $n$ is odd, append one zero column; if $n$ is even, leave the matrix unchanged. Let $C$ be the row span of this even-length matrix. The forward construction produces the unital triorthogonal space
\[
\mathcal U=C+\operatorname{span}\{\mathbf1\}.
\]
If $\mathbf1\in C$, then $\mathcal U=C$, and retaining all rows recovers the stabilisers up to a basis change. Otherwise, $\mathbf1$ was added to $C$ to form $\mathcal{U}$. A given generator matrix for a unital triorthogonal space could have been formed from $G_0$ by adding the top all-ones row to any combination of the lower $m$ rows, where this choice is encoded in the column vector $b \in \mathbb{F}_2^m$. Our procedure reverses all these possible choices.

Different choices may produce equivalent stabiliser matrices, which we identify under row operations and column permutations. Since $c$ is even, even protocol length $n$ uses a space of length $c=n$, while odd $n$ uses $c=n+1$. Thus all required spaces for $n\leq54$ have length at most $54$. In both constructions, logical rows remain to be added, as described next.

\subsection{Adding compatible logical rows}

We now suppose we have some stabiliser rows given by a matrix $G_0$. This is some distinct set of nonzero columns $X \subseteq \mathbb{F}_2^r \setminus \{0\}$ spanning $\mathbb{F}_2^r$ (since we assume $G_0$ has full row rank) and satisfying Equation~\eqref{eq:moments}; we ask which logical rows may be added. This construction extends and formalises the compatible-row approach appearing in Refs.~\cite{haah2018codes,gong2026extension}. A row is a function $\ell:X\rightarrow\F_2$, whose entry at the column indexed by $u\in X$ is $\ell(u)$. The overlap conditions involving one logical row define the linear space
\begin{equation}
V_X=\left\{\ell:X\to\F_2:\sum_{u\in X}\ell(u)p(u)=0
\text{ for every square-free monomial $p$ of degree $1$ or $2$}\right\}.
\label{eq:VX}
\end{equation}
Let $L_X$ be the restrictions to $X$ of linear functions $u\mapsto\alpha\cdot u$, with $\alpha\in\F_2^r$. This is exactly $\operatorname{rowspan}(G_0)$, viewed as functions on $X$, and Equation~\eqref{eq:moments} gives $L_X\subseteq V_X$. Adding an element of $L_X$ to a logical row adds a sum of stabiliser rows; we want to identify two logical rows if they differ by a sum of stabiliser rows, and so we work in $V_X/L_X$.

It remains to impose the overlap conditions involving two logical rows. For each stabiliser coordinate $a$, define
\begin{equation}
B_a(\ell,\ell')=\sum_{u\in X}\ell(u)\ell'(u)u_a.
\label{eq:Ba}
\end{equation}
These are symmetric bilinear forms. We also have $B_a(\ell,\ell)=\sum_{u\in X}\ell(u)u_a=0$. They are well-defined on the quotient: adding a linear function to either argument gives a vanishing difference using Equation~\eqref{eq:VX}.

A subspace $U\subseteq V_X/L_X$ is \emph{totally isotropic with respect to these forms} when $B_a(\ell,\ell')=0$ for all $\ell,\ell'\in U$ and every $a$. Notice that this is a condition on the whole subspace, not just on its individual vectors. Bilinearity means that it is enough to check pairs over any basis of $U$. Isotropy is exactly the remaining overlap conditions involving two logical rows and one stabiliser row.

Choose representatives $\ell_1,\ldots,\ell_k$ of a basis of such a subspace and form the columns of a generalised triorthogonal matrix
\begin{equation}
g_u=(\ell_1(u),\ldots,\ell_k(u),u)^T,\qquad u\in X.
\label{eq:reverse-columns}
\end{equation}
The columns are distinct and nonzero because their stabiliser parts are distinct and nonzero. Equations~\eqref{eq:VX} and \eqref{eq:Ba} give all overlaps involving stabiliser rows. Full row rank follows from the span of $X$ and independence modulo $L_X$. Finally, nonzero stabiliser columns exclude weight-one logical errors, and distinct stabiliser columns exclude weight-two logical errors. Thus every such construction has $d_Z\geq3$. On an allowed $U$, we may write the symmetric trilinear form describing the logical action as $\tau(\ell,\ell',\ell'')=\sum_{u\in X}\ell(u)\ell'(u)\ell''(u)$. This does not depend on the representatives modulo $L_X$: changing one representative by a stabiliser function adds a sum of bilinear forms $B_a(\ell',\ell'')$, which vanish on $U$.

Note that in practice, for our classification, we add logical rows to candidate stabiliser spaces one at a time, in order to find all sets of logical rows that may be added. Each additional logical row adds more linear constraints on the remaining logical rows that may be added. See the repository~\cite{repo} for implementation details.

The forward and reverse constructions show that every generalised triorthogonal matrix in scope is obtained. Notice that from a length $c$ unital triorthogonal space, we obtain generalised triorthogonal codes with length $n=c$ or $n=c-1$. Note that Nezami and Haah use a puncturing-based correspondence between triorthogonal matrices and unital triorthogonal spaces. The result is that their classification of unital triorthogonal spaces up to length $38$ only yields a classification of triorthogonal codes with $n$/$k$ input/output $\ket{\mathrm{T}}$ states with $n+k \leq 38$. We comment that a puncturing-based correspondence is possible to classify generalised triorthogonal matrices from unital triorthogonal spaces, but the adding of logical rows yields a longer generalised triorthogonal code classification from the same unital triorthogonal space classification.

\subsection{Additional considerations on distance and space footprint}

We make additional remarks here on distance and space footprint which can constrain our search space in adding logical rows. Write $\lambda(u)=(\ell_1(u),\ldots,\ell_k(u))$ for the logical part of a column. An error with support $E\subseteq X$ has zero stabiliser syndrome when $\sum_{u\in E}u=0$, and it is a logical error when $\sum_{u\in E}\lambda(u)\neq0$. Thus
\begin{equation}
d_Z=\min\left\{|E|:E\subseteq X,\ \sum_{u\in E}u=0,\ \sum_{u\in E}\lambda(u)\neq0\right\}.
\end{equation}
For $d_Z=3$, these are precisely the unordered triples of distinct points $u,v,w\in X$ satisfying $u+v+w=0$ and
\begin{equation}
\lambda(u)+\lambda(v)+\lambda(w)\neq0.
\end{equation}
Their number is the leading error coefficient for independent input errors of probability $p$, so the conditional output error is $a_3p^3+O(p^4)$. More generally, the coefficient $a_d$ counts the weight-$d$ logical errors when $d_Z=d$.

Enforcing a distance lower bound in fact allows us to constrain the addition of logical rows. For every set $E\subseteq X$ with $|E|<d_0$ and $\sum_{u\in E}u=0$, require $\sum_{u\in E}\ell(u)=0$ for every candidate row $\ell$. These are linear constraints, and all stabiliser functions satisfy them. They therefore define a subspace of the same quotient $V_X/L_X$, within which we impose the same bilinear overlap conditions to search for compatible logical rows. The resulting matrices have $d_Z\geq d_0$; we later calculate the exact distance of the final protocol.

The distinct, nonzero stabiliser columns also give $n\leq2^r-1$, where $r=N-k$. In our scope, this gives the bound
\begin{equation}
N\geq k+\lceil\log_2(n+1)\rceil.
\end{equation}

\section{Unital triorthogonal space classification using directional derivatives}

\subsection{The Reed-Muller correspondence}

We now describe how we classify the unital triorthogonal spaces used above. We use the same correspondence established by Nezami and Haah, described shortly, that the unital triorthogonal spaces of length $c$ and dimension $m+1$ are in one-to-one correspondence with the codewords of Reed-Muller codes $\RM(m-4,m)$ of weight $c$, up to affine equivalence (defined shortly). We comment that the classification of Reed-Muller codewords of certain weights up to affine equivalence is a classical problem, and Nezami and Haah use classical results to classify unital triorthogonal spaces up to length $38$~\cite{kasami1970weight,kasami1976weight}. We use some tricks with directional derivatives inspired by more recent literature on asymptotic bounds of weight distributions of Reed-Muller codes~\cite{kaufman2012weight,abbe2015reed,sberlo2020performance} to push the classification to length $54$.

Write $c$ for the length of a unital triorthogonal space, to distinguish it from the protocol length $n$, and $m+1$ for its dimension. Choosing the all-ones vector as the first generator row gives distinct columns $(1,y)^T$, with $y$ running over a set $Y\subseteq\F_2^m$ of size $c$. The generator matrix for the space must have $m+1$ independent rows, the first row of which is the all-ones row.

Every function $f:\F_2^m \to \F_2$ has a unique square-free polynomial expression over $\mathbb{F}_2$, with $x_i^2=x_i$. The degree of the function is the largest degree of a monomial in this expression. The binary Reed-Muller code $\RM(\rho,m)$ consists of the evaluations of such polynomials of degree at most $\rho$. Note that the evaluations of the functions (polynomials) over $\F_2^m$ are in one-to-one correspondence with the functions themselves, and so we identify them. Any function $f:\F_2^m \to \F_2$ is also in one-to-one correspondence with its support (the set of points at which it evaluates to $1$). Now, to a generator matrix for a unital triorthogonal space with columns $(1,y)^T$, for $y \in Y \subseteq \F_2^m$, we associate the function $f$ which is exactly supported on $Y$. This is also called the indicator function on $Y$; we write $f = 1_Y$. The fact that the space is unital triorthogonal is the same as saying
\begin{equation}
    0 = \sum_{u \in Y}p(u) = \sum_{u \in \F_2^m}1_Y(u)p(u)
\end{equation}
for every square-free monomial $p$ of degree $0$, $1$, $2$ or $3$, which is the same as saying that $1_Y \in \RM(m-4,m)$, because $\RM(m-4,m)$ is the dual of $\RM(3,m)$.

Consider performing row operations on a generator matrix for a unital triorthogonal space, where we always preserve the all-ones row at the top. The result is a generator matrix for the same space. One can also consider performing an \textit{affine transformation} on the variables of any function $f:\F_2^m \to \F_2$ (such as the examples we care about $f = 1_Y \in \RM(m-4,m)$) via $y \mapsto Ay + b$, where $A$ is an invertible $m \times m$ binary matrix, and $b \in \F_2^m$. We note that by the above matrix-to-function correspondence, these two transformations are the same. In conclusion, to classify unital triorthogonal spaces of length $c$, we must classify Reed-Muller codewords in $\RM(m-4,m)$ of weight $c$, up to this affine equivalence. Before moving on, we comment that one can check that if we want to guarantee that the unital triorthogonal space arising from a generator matrix with $1+m$ rows has dimension $1+m$ and not less (i.e. there is no dependence amongst their rows), this exactly means that the corresponding function $f = 1_Y$ is not divisible by some nonconstant degree-one polynomial.

\subsection{Choosing a directional derivative with few double pairs}

Let $f:\F_2^m \to \F_2$ have degree at most $m-4$, whose support we denote by $Y$, which has size $c$, as above. For a nonzero $a\in\F_2^m$, define the derivative of $f$ in the direction $a$ as $\Delta_af(x)=f(x)+f(x+a)$. Note that the pairs $\{x,x+a\}$ partition the space $\F_2^m$. Let $\nu_a$ count those pairs whose two points both belong to $Y$. Every unordered pair of distinct points in $Y$ has exactly one nonzero difference, and hence
\begin{equation}
\sum_{a\neq0}\nu_a=\binom{c}{2}.
\label{eq:fibre-average}
\end{equation}
There is therefore a direction for which
\begin{equation}
\nu_a\leq\left\lfloor\frac{\binom{c}{2}}{2^m-1}\right\rfloor.
\label{eq:double-pair-bound}
\end{equation}
In words, we are always guaranteed the existence of an $a$ for which $\nu_a$ is quite small for the cases we care about.

After an affine change of coordinates, write this direction as the last coordinate direction, and put $z=(x_1,\ldots,x_{m-1})$, $t=x_m$. As in the main text, write
\begin{align}
f(z,t)&=f_0(z)+t\,g(z),\\
f_0(z)&=f(z,0),\qquad f_1(z)=f(z,1),\qquad g=f_0+f_1.
\end{align}
Then $\deg f_0\leq m-4$ and $\deg g\leq m-5$. In particular, $g\in\RM(m-5,m-1)$ is an indicator polynomial for a unital support in fewer variables. Define
\begin{align}
P&=\{z\in\F_2^{m-1}:f(z,0)\neq f(z,1)\}=\operatorname{supp}(g),\\
D&=\{z\in\F_2^{m-1}:f(z,0)=f(z,1)=1\}.
\end{align}
We have
\begin{equation}
c=|P|+2|D|,\qquad |D|=\nu_a,\qquad \wt(g)=c-2\nu_a,
\end{equation}
where $|D| = \nu_a$ follows by definition of the direction $a$. Note that we have chosen the direction $a$ to maximise the weight of the derivative, which is $2\wt(g)$, which is equivalent to minimising $|D|$.

Note that we need not actually find the optimal direction $a$ in our classification. We may simply imagine the affine transformation having been performed, and consider every $\nu_a$ satisfying the bound of Equation~\eqref{eq:double-pair-bound}. This will become more clear when we consider reconstructing the polynomial $f$ from the derivative below.

\subsection{Reconstructing the original polynomial}

Our actual classification is performed recursively, constructing catalogues from lower-variable catalogues, effectively reversing the above procedure. Indeed, for every $\nu$ satisfying Equation~\eqref{eq:double-pair-bound}, imagine fixing some $g$ of weight $c-2\nu$ from some lower-variable catalogue, and put $P=\operatorname{supp}(g)$ as above. We aim to reconstruct all possible $f$; the resulting $f_0$ gets constructed as part of the reconstruction. Consider every choice of set
\begin{equation}
D\subseteq\F_2^{m-1}\setminus P,\qquad |D|=\nu.
\end{equation}
Notice that we upper bounded the possible $\nu$ to make this as computationally tractable as possible. The only remaining choice to make to specify $f$'s entire support is to let $b_z = f_0(z)$, for each $z \in P$, be an unknown binary variable. When $b_z = 1$, it tells us that $f$ is supported on $(z,0)$, and not $(z,1)$, whereas when $b_z=0$, it tells us that $f$ is supported on $(z,1)$, and not $(z,0)$.

The choices for the variables $b_z$ are reduced by enforcing the condition $\deg(f_0)\leq m-4$. $\RM(m-4,m-1)$ is the dual of $\RM(2,m-1)$. The condition $\deg(f_0)\leq m-4$ is thus exactly
\begin{equation}
\sum_{z\in P}b_zp(z)=\sum_{z\in D}p(z)
\quad\text{for every square-free monomial $p$ of degree at most $2$, including $1$.}
\label{eq:lift-linear-system}
\end{equation}
For fixed $P,D$, this is a linear system over $\F_2$ in the variables $b_z$. Some choices of $D$ give no solution and are discarded; every obtained solution gives a polynomial $f=f_0+t g$ of the required degree and weight. Conversely, every admissible $f$ determines one of these choices and solutions. The recursion may be used to perform the classification. Note that the recursion decreases the number of variables, but it also decreases the weight when $\nu>0$. In the repository~\cite{repo}, we give further tricks for decreasing the search space, including a fast method for determining when a given choice of the set $D$ can yield a solution to the system in Equation~\eqref{eq:lift-linear-system}. As another example, it is useful to upper bound the dimension $m+1$ of a unital triorthogonal space of length $c$ for our classification. If the unital triorthogonal space is $H$, we have $H \subseteq H^\perp$, implying $m \leq \left\lfloor\frac{c}{2}\right\rfloor -1$. However, we show in~\cite{repo} that the stronger bound
\begin{equation}
    m \leq \left\lfloor\frac{c+\left\lfloor c/16\right\rfloor}{3}\right\rfloor -1
\end{equation}
holds, also reducing the search space.
\subsection{The resulting space catalogues}

We comment that the unital triorthogonal space classification of Nezami and Haah~\cite{nezami2022classification} up to length $38$ omits two spaces of dimension $10$ at length $36$. The full classification up to length $38$ contains $40$ spaces. More generally, the number of unital triorthogonal spaces at each length up to $54$ is shown in Table~\ref{tab:space-counts}; the full catalogues are supplied in the accompanying data~\cite{classificationdata}. As one might expect, the number of unital triorthogonal spaces appears to increase exponentially as the length $c$ grows, likely rendering significant further enumeration of unital triorthogonal spaces impractical without new ideas.

\begin{table}[htb]
\centering
\caption{Numbers of unital triorthogonal spaces of length $c$, up to coordinate permutations, with distinct generator columns. There are no such spaces at odd lengths or at lengths below $16$.}
\label{tab:space-counts}
\begin{tabular}{@{}r@{\hspace{1.5em}}r@{\hspace{5em}}r@{\hspace{1.5em}}r@{}}
\toprule
Length $c$ & Number of spaces & Length $c$ & Number of spaces \\
\midrule
16 & $1$  & 36 & $16$ \\
18 & $0$  & 38 & $8$ \\
20 & $0$  & 40 & $110$ \\
22 & $0$  & 42 & $86$ \\
24 & $1$  & 44 & $863$ \\
26 & $0$  & 46 & $2\,403$ \\
28 & $2$  & 48 & $28\,875$ \\
30 & $1$  & 50 & $304\,611$ \\
32 & $10$ & 52 & $7\,519\,688$ \\
34 & $1$  & 54 & $293\,172\,583$ \\
\bottomrule
\end{tabular}
\end{table}

\section{The Pareto Frontier}

For a fixed output class and exact distance, we call a protocol Pareto optimal if no other protocol is at least as good in both $(n,N)$ and strictly better in one. This is the definition used in the main text. Different distances and different output classes are separate comparisons.

\begin{samepage}
The final frontier contains the following points:
\begin{center}
\setlength{\tabcolsep}{7pt}
\begin{tabular}{r@{\quad}r@{\quad}r@{\quad}r@{\quad}l@{\quad}l}
\toprule
$k$ & output classes & frontier points & distances & $n$ range & $N$ range\\
\midrule
1 & 1 & 5 & $3,4,5$ & $15$--$53$ & $5$--$12$\\
2 & 2 & 4 & $3,4$ & $28$--$48$ & $8$--$10$\\
3 & 6 & 9 & $3,4$ & $35$--$48$ & $9$--$10$\\
4 & 13 & 15 & $3$ & $43$--$47$ & $10$--$11$\\
5 & 28 & 29 & $3,4$ & $47$--$52$ & $11$--$12$\\
6 & 10 & 10 & $3$ & $47$--$52$ & $12$--$13$\\
7 & 2 & 2 & $3$ & $48$ & $13$\\
\midrule
total & 62 & 74 & & &\\
\bottomrule
\end{tabular}
\end{center}
\end{samepage}
As stated in the main text, there are 74 points spanning 62 output classes: 67 have distance three, five have distance four, and two have distance five. Table~\ref{tab:complete-pareto} gives every point and its representative gate, and the accompanying data~\cite{classificationdata} give the corresponding matrices and error coefficients. This protocol catalogue is also included in the repository~\cite{repo}.

\subsection{Literature comparison}

There are $74$ Pareto points for $62$ output classes at lengths $n\leq54$ and distances $d_Z\geq3$. Our literature comparison identifies $65$ of these points as new and the remaining nine as existing. We compare points at fixed $\mathrm{CNOT}+\mathrm{S}$ output class and exact distance, using the length $n$ and spatial footprint $N$, where $N$ counts all matrix rows. We do not count a point as new if a published protocol, or a simple modification of one, attains the same output and distance with no greater $n$ or $N$. Such modifications include changes of logical basis, restrictions to fewer logical rows, and conversion of $\mathrm{CS}$ or $\mathrm{CCZ}$ inputs into $\mathrm{T}$ inputs. The nine existing protocols are explained below.

Five points occur explicitly in published work. Bravyi and Kitaev's $15\mathrm{T}\to \mathrm{T}$ protocol has $(n,N,d_Z)=(15,5,3)$, with one logical row and four stabiliser rows~\cite{bravyi2005universal}. Nezami and Haah give $28\mathrm{T}\to2\mathrm{T}$ with $(n,N,d_Z)=(28,9,3)$, and $35\mathrm{T}\to3\mathrm{T}$ with $(n,N,d_Z)=(35,9,3)$~\cite{nezami2022classification}. Their matrices have seven and six stabiliser rows, respectively. Jacinto et al.\ give $47\mathrm{T}\to\CCZ$ with $(n,N,d_Z)=(47,9,3)$ and $48\mathrm{T}\to\CCZ$ with $(n,N,d_Z)=(48,10,4)$~\cite{jacinto2026compact}. These match our output classes and resource parameters, even when the chosen matrix or leading error coefficient differs.

Two further points are obtained by restricting the logical rows of Nezami and Haah's $35\mathrm{T}\to3\mathrm{T}$ matrix~\cite{nezami2022classification}. Write its three logical rows as $g_1,g_2,g_3$, and retain all six stabiliser rows. Keeping $g_1,g_2$ gives $35\mathrm{T}\to2\mathrm{T}$ with $(n,N,d_Z)=(35,8,3)$. Instead keeping $g_1+g_3,g_2+g_3$ gives two even-weight logical rows with odd pair overlap, and hence $35\mathrm{T}\to\CS$ with the same $(n,N,d_Z)=(35,8,3)$. Restricting logical rows cannot decrease the distance, and we check directly that both distances remain exactly three. They are marked with $\dagger$ in Table~\ref{tab:complete-pareto}.

The final two existing points are simple modifications of the binarised matrices of Gong et al.~\cite{gong2026extension}. We use their Construction 4.21, which evaluates the logical rows $xy,x^2,y^2$ and stabiliser rows $x,y,1$ at all 16 points of $\F_4^2$. Before the final logical fixing used there to obtain $\mathrm{TOF}\#$, this construction has six binary logical rows and six binary stabiliser rows. Retaining all six logical rows gives the starting point for both protocols below.

\phantomsection
\label{sec:field-conversion}
First perform binarisation in the ordered trace-self-dual basis $(\beta_1,\beta_2)=(\omega,\omega^2)$, where $\omega^2+\omega+1=0$. Each field entry $\gamma$ is replaced by the $2\times2$ binary matrix with entries $\operatorname{Tr}_{\F_4/\F_2}(\gamma\beta_a\beta_b)$. In the field-row order above, this gives a $12\times32$ binary matrix $B$, with each consecutive pair of columns corresponding to one $\mathrm{CS}$ input. Next apply the $\mathrm{CS}$-to-$\mathrm{T}$ embedding described in Ref.~\cite{gong2026extension}: for each such pair of column vectors $a_j,b_j\in\F_2^{12}$, make the replacement
\begin{equation}
(a_j,b_j)\longmapsto(a_j,b_j,a_j+b_j),
\qquad j=1,\ldots,16.
\end{equation}
This simply appends the parity column for each pair. It implements the three-$\mathrm{T}$ synthesis of $\mathrm{CS}$, using $\mathrm{T},\mathrm{T},\mathrm{T}^\dagger$ on the three columns; $\mathrm{T}^\dagger$ differs from $\mathrm{T}$ by an $\mathrm{S}$ correction. No matrix rows are added, so the resulting generalised triorthogonal matrix $G$ has $n=48$ and $N=12$.

This embedding preserves exact block $\mathrm{Z}$-error distance. The three columns represent the three nonzero error patterns on the original two-qubit block. Any error in the embedded matrix induces the same stabiliser and logical syndromes as a block error affecting no more blocks than its weight. Conversely, every nonzero block error is realised by a single error on its corresponding parity column. The minimum undetected logical-error weight is therefore unchanged, giving $d_Z=3$. An invertible change of logical basis, together with diagonal Clifford corrections, puts the six-qubit output in the displayed $\mathrm{CNOT}+\mathrm{S}$ representative
\begin{equation}
U_6=\FieldOutputSix.
\end{equation}
Thus retaining the logical rows before Gong et al.'s final fixing, followed by the $\mathrm{CS}$-to-$\mathrm{T}$ embedding and a logical basis change, gives the point $48\mathrm{T}\to U_6|+\rangle^{\otimes6}$ with $(n,N,d_Z)=(48,12,3)$.

For the five-qubit point, apply a logical restriction; delete the third logical row of $G$, in the original binary row order specified above and before the basis change used to display $U_6$. Equivalently, fix that logical bit to zero. All six stabiliser rows and all 48 columns are retained, leaving five logical rows and $N=11$. Row deletion preserves the required overlap conditions and cannot decrease the distance; direct checking gives exact distance three. A change of logical basis and diagonal Clifford corrections give
\begin{equation}
U_5=\FieldOutputFive,
\end{equation}
so this restriction yields $48\mathrm{T}\to U_5|+\rangle^{\otimes5}$ with $(n,N,d_Z)=(48,11,3)$. This five-qubit output is not explicitly named in Ref.~\cite{gong2026extension}, but follows by the stated row deletion from its binarised construction. Both points are therefore counted as existing, and marked with $*$ in Table~\ref{tab:complete-pareto} to indicate these derivations.

\begin{center}
\begin{tabular}{lrrr}
\toprule
protocol & $n$ & $N$ & $d_Z$\\
\midrule
$48\mathrm{T}\to U_6|+\rangle^{\otimes6}$ & 48 & 12 & 3\\
$48\mathrm{T}\to U_5|+\rangle^{\otimes5}$ & 48 & 11 & 3\\
\bottomrule
\end{tabular}
\end{center}

\clearpage

\subsection{Catalogue of Protocols}
\begingroup
\small
\setlength{\tabcolsep}{4pt}
\renewcommand{\arraystretch}{1.03}
\setlength{\LTpre}{4pt}
\setlength{\LTpost}{6pt}
\setlength{\LTcapwidth}{\textwidth}
\begin{longtable}{@{}c>{\raggedright\arraybackslash}p{0.67\textwidth}rrrc@{}}
\caption{All 74 optimal generalised triorthogonal protocols, with $n\leq54$ and distance $d_Z\geq3$. Each row describes $n\mathrm{T}\to U|+\rangle^{\otimes k}$; $N$ is the spatial footprint, namely the number of rows of the matrix describing the protocol. Outputs are identified up to $\mathrm{CNOT}+\mathrm{S}$ equivalence, and a protocol is discarded if it is worse than another with the same output in its input $\ket{\mathrm{T}}$ count $n$, spatial footprint $N$, and distance $d_Z$. References identify published protocols, or those that may be easily derived from published protocols. A dash means that the protocol is found to be new in the literature. $\dagger$: the protocol may be obtained as a simple modification of the $35 \mathrm{T} \to 3\mathrm{T}$ protocol with distance $3$ from~\cite{nezami2022classification}, see above. $*$: the protocol may be obtained as a simple modification of a matrix in~\cite{gong2026extension}, see above.}\label{tab:complete-pareto}\\
\toprule
$k$ & Representative output gate $U$ & $d_Z$ & $n$ & $N$ & Reference\\
\midrule
\endfirsthead
\multicolumn{6}{l}{\textbf{Table \thetable.} Continued.}\\
\toprule
$k$ & Representative gate $U$ & $d_Z$ & $n$ & $N$ & Reference\\
\midrule
\endhead
\midrule
\multicolumn{6}{r}{Continued on next page.}\\
\endfoot
\bottomrule
\endlastfoot
1 & $\mathrm{T}$ & $3$ & 15 & 5 & \cite{bravyi2005universal} \\
\noalign{\penalty10000}
 &  & $4$ & 52 & 10 & -- \\
\noalign{\penalty10000}
 &  & $4$ & 48 & 11 & -- \\
\noalign{\penalty10000}
 &  & $5$ & 53 & 11 & -- \\
\noalign{\penalty10000}
 &  & $5$ & 49 & 12 & -- \\
\addlinespace[3pt]
2 & $\mathrm{T}^{\otimes 2}$ & $3$ & 35 & 8 & \cite{nezami2022classification}$^{\dagger}$ \\
\noalign{\penalty10000}
 &  & $3$ & 28 & 9 & \cite{nezami2022classification} \\
2 & $\mathrm{CS}$ & $3$ & 35 & 8 & \cite{nezami2022classification}$^{\dagger}$ \\
\noalign{\penalty10000}
 &  & $4$ & 48 & 10 & -- \\
\addlinespace[3pt]
3 & $\mathrm{T}^{\otimes 3}$ & $3$ & 35 & 9 & \cite{nezami2022classification} \\
3 & $\mathrm{T}\otimes\allowbreak \mathrm{CS}$ & $3$ & 47 & 9 & -- \\
\noalign{\penalty10000}
 &  & $3$ & 43 & 10 & -- \\
3 & $\mathrm{CS}_{12}\allowbreak\mathrm{CS}_{13}$ & $3$ & 36 & 9 & -- \\
3 & $\mathrm{T}_{1}\allowbreak\mathrm{CS}_{12}\allowbreak\mathrm{CS}_{13}$ & $3$ & 47 & 9 & -- \\
\noalign{\penalty10000}
 &  & $3$ & 43 & 10 & -- \\
3 & $\mathrm{T}_{1}\allowbreak\mathrm{CCZ}_{123}$ & $3$ & 47 & 9 & -- \\
3 & $\mathrm{CCZ}$ & $3$ & 47 & 9 & \cite{jacinto2026compact} \\
\noalign{\penalty10000}
 &  & $4$ & 48 & 10 & \cite{jacinto2026compact} \\
\addlinespace[3pt]
4 & $\mathrm{T}^{\otimes 4}$ & $3$ & 47 & 10 & -- \\
\noalign{\penalty10000}
 &  & $3$ & 44 & 11 & -- \\
4 & $\mathrm{T}^{\otimes 2}\otimes\allowbreak \mathrm{CS}$ & $3$ & 47 & 10 & -- \\
4 & $\mathrm{T}\otimes\allowbreak \bigl(\mathrm{CS}_{12}\allowbreak\mathrm{CS}_{13}\bigr)$ & $3$ & 47 & 10 & -- \\
\noalign{\penalty10000}
 &  & $3$ & 43 & 11 & -- \\
4 & $\mathrm{T}\otimes\allowbreak \bigl(\mathrm{T}_{1}\allowbreak\mathrm{CS}_{12}\allowbreak\mathrm{CS}_{13}\bigr)$ & $3$ & 47 & 10 & -- \\
4 & $\mathrm{T}\otimes\allowbreak \bigl(\mathrm{T}_{1}\allowbreak\mathrm{CCZ}_{123}\bigr)$ & $3$ & 47 & 10 & -- \\
4 & $\mathrm{T}\otimes\allowbreak \mathrm{CCZ}$ & $3$ & 47 & 10 & -- \\
4 & $\mathrm{T}_{1}\allowbreak\mathrm{CS}_{12}\allowbreak\mathrm{CCZ}_{134}$ & $3$ & 47 & 10 & -- \\
4 & $\mathrm{CS}_{12}\allowbreak\mathrm{CCZ}_{134}$ & $3$ & 47 & 10 & -- \\
4 & $\mathrm{CS}^{\otimes 2}$ & $3$ & 47 & 10 & -- \\
4 & $\mathrm{CS}_{12}\allowbreak\mathrm{CS}_{13}\allowbreak\mathrm{CS}_{24}$ & $3$ & 47 & 10 & -- \\
4 & $\mathrm{T}_{1}\allowbreak\mathrm{CS}_{12}\allowbreak\mathrm{CS}_{13}\allowbreak\mathrm{CS}_{24}$ & $3$ & 47 & 10 & -- \\
4 & $\mathrm{T}_{1}\allowbreak\mathrm{T}_{2}\allowbreak\mathrm{CS}_{12}\allowbreak\mathrm{CS}_{13}\allowbreak\mathrm{CS}_{24}$ & $3$ & 47 & 10 & -- \\
4 & $\mathrm{T}_{1}\allowbreak\mathrm{CS}_{23}\allowbreak\mathrm{CCZ}_{124}$ & $3$ & 47 & 10 & -- \\
\addlinespace[3pt]
5 & $\mathrm{T}\otimes\allowbreak \mathrm{CS}^{\otimes 2}$ & $3$ & 52 & 12 & -- \\
5 & $\mathrm{T}\otimes\allowbreak \bigl(\mathrm{CS}_{12}\allowbreak\mathrm{CS}_{13}\allowbreak\mathrm{CS}_{24}\bigr)$ & $3$ & 52 & 12 & -- \\
5 & $\mathrm{CS}\otimes\allowbreak \mathrm{CCZ}$ & $3$ & 51 & 12 & -- \\
5 & $\mathrm{T}^{\otimes 2}\otimes\allowbreak \mathrm{CCZ}$ & $3$ & 51 & 12 & -- \\
5 & $\mathrm{T}^{\otimes 2}\otimes\allowbreak \bigl(\mathrm{CS}_{12}\allowbreak\mathrm{CS}_{13}\bigr)$ & $3$ & 48 & 12 & -- \\
5 & $\mathrm{CS}_{15}\allowbreak\mathrm{CS}_{34}\allowbreak\mathrm{CCZ}_{125}\allowbreak\mathrm{CCZ}_{234}$ & $3$ & 48 & 11 & -- \\
5 & $\mathrm{CS}\otimes\allowbreak \bigl(\mathrm{CS}_{12}\allowbreak\mathrm{CS}_{13}\bigr)$ & $3$ & 51 & 12 & -- \\
5 & $\mathrm{CS}_{45}\allowbreak\mathrm{CCZ}_{135}\allowbreak\mathrm{CCZ}_{234}$ & $3$ & 48 & 11 & \cite{gong2026extension}$^{*}$ \\
5 & $\mathrm{T}_{5}\allowbreak\mathrm{CCZ}_{145}\allowbreak\mathrm{CCZ}_{234}$ & $3$ & 47 & 11 & -- \\
5 & $\mathrm{CS}_{25}\allowbreak\mathrm{CS}_{34}\allowbreak\mathrm{CCZ}_{145}$ & $3$ & 48 & 11 & -- \\
5 & $\mathrm{T}_{5}\allowbreak\mathrm{CS}_{25}\allowbreak\mathrm{CS}_{34}\allowbreak\mathrm{CCZ}_{145}\allowbreak\mathrm{CCZ}_{234}$ & $3$ & 52 & 12 & -- \\
5 & $\mathrm{T}^{\otimes 3}\otimes\allowbreak \mathrm{CS}$ & $3$ & 51 & 12 & -- \\
5 & $\mathrm{T}_{5}\allowbreak\mathrm{CS}_{14}\allowbreak\mathrm{CS}_{23}\allowbreak\mathrm{CCZ}_{234}\allowbreak\mathrm{CCZ}_{235}$ & $3$ & 52 & 12 & -- \\
5 & $\mathrm{CCZ}_{145}\allowbreak\mathrm{CCZ}_{235}$ & $3$ & 47 & 11 & -- \\
\noalign{\penalty10000}
 &  & $4$ & 48 & 12 & -- \\
5 & $\mathrm{T}_{5}\allowbreak\mathrm{CCZ}_{145}\allowbreak\mathrm{CCZ}_{235}$ & $3$ & 47 & 11 & -- \\
5 & $\mathrm{T}_{5}\allowbreak\mathrm{CS}_{14}\allowbreak\mathrm{CS}_{23}\allowbreak\mathrm{CCZ}_{145}\allowbreak\mathrm{CCZ}_{235}$ & $3$ & 47 & 11 & -- \\
5 & $\mathrm{T}^{\otimes 5}$ & $3$ & 47 & 11 & -- \\
5 & $\mathrm{CS}_{25}\allowbreak\mathrm{CS}_{34}\allowbreak\mathrm{CCZ}_{145}\allowbreak\mathrm{CCZ}_{234}\allowbreak\mathrm{CCZ}_{235}$ & $3$ & 48 & 11 & -- \\
5 & $\mathrm{CS}_{45}\allowbreak\mathrm{CCZ}_{145}\allowbreak\mathrm{CCZ}_{235}$ & $3$ & 48 & 11 & -- \\
5 & $\mathrm{CS}_{12}\allowbreak\mathrm{CS}_{34}\allowbreak\mathrm{CCZ}_{135}\allowbreak\mathrm{CCZ}_{245}$ & $3$ & 48 & 11 & -- \\
5 & $\mathrm{T}_{5}\allowbreak\mathrm{CS}_{14}\allowbreak\mathrm{CS}_{23}\allowbreak\mathrm{CCZ}_{345}$ & $3$ & 52 & 12 & -- \\
5 & $\mathrm{T}\otimes\allowbreak \bigl(\mathrm{T}_{1}\allowbreak\mathrm{CS}_{12}\allowbreak\mathrm{CCZ}_{134}\bigr)$ & $3$ & 47 & 11 & -- \\
5 & $\mathrm{T}_{5}\allowbreak\mathrm{CS}_{34}\allowbreak\mathrm{CCZ}_{125}\allowbreak\mathrm{CCZ}_{345}$ & $3$ & 51 & 12 & -- \\
5 & $\mathrm{T}^{\otimes 2}\otimes\allowbreak \bigl(\mathrm{T}_{1}\allowbreak\mathrm{CCZ}_{123}\bigr)$ & $3$ & 51 & 12 & -- \\
5 & $\mathrm{T}_{5}\allowbreak\mathrm{CS}_{25}\allowbreak\mathrm{CS}_{34}\allowbreak\mathrm{CCZ}_{125}\allowbreak\mathrm{CCZ}_{345}$ & $3$ & 51 & 12 & -- \\
5 & $\mathrm{T}\otimes\allowbreak \bigl(\mathrm{T}_{1}\allowbreak\mathrm{CS}_{12}\allowbreak\mathrm{CS}_{13}\allowbreak\mathrm{CS}_{24}\bigr)$ & $3$ & 51 & 12 & -- \\
5 & $\mathrm{T}^{\otimes 2}\otimes\allowbreak \bigl(\mathrm{T}_{1}\allowbreak\mathrm{CS}_{12}\allowbreak\mathrm{CS}_{13}\bigr)$ & $3$ & 51 & 12 & -- \\
5 & $\mathrm{T}\otimes\allowbreak \bigl(\mathrm{T}_{1}\allowbreak\mathrm{T}_{2}\allowbreak\mathrm{CS}_{12}\allowbreak\mathrm{CS}_{13}\allowbreak\mathrm{CS}_{24}\bigr)$ & $3$ & 52 & 12 & -- \\
\addlinespace[3pt]
6 & $\mathrm{T}\otimes\allowbreak \bigl(\mathrm{T}_{5}\allowbreak\mathrm{CS}_{14}\allowbreak\mathrm{CS}_{23}\allowbreak\mathrm{CCZ}_{145}\allowbreak\mathrm{CCZ}_{235}\bigr)$ & $3$ & 52 & 13 & -- \\
6 & $\mathrm{CS}_{12}\allowbreak\mathrm{CS}_{34}\allowbreak\mathrm{CCZ}_{125}\allowbreak\mathrm{CCZ}_{135}\allowbreak\mathrm{CCZ}_{136}\allowbreak\mathrm{CCZ}_{246}\allowbreak\mathrm{CCZ}_{345}$ & $3$ & 48 & 12 & -- \\
6 & $\mathrm{CS}_{12}\allowbreak\mathrm{CCZ}_{134}\allowbreak\mathrm{CCZ}_{135}\allowbreak\mathrm{CCZ}_{146}\allowbreak\mathrm{CCZ}_{234}\allowbreak\mathrm{CCZ}_{256}$ & $3$ & 48 & 12 & \cite{gong2026extension}$^{*}$ \\
6 & $\mathrm{T}^{\otimes 3}\otimes\allowbreak \bigl(\mathrm{CS}_{12}\allowbreak\mathrm{CS}_{13}\bigr)$ & $3$ & 51 & 13 & -- \\
6 & $\mathrm{CS}_{12}\allowbreak\mathrm{CS}_{34}\allowbreak\mathrm{CCZ}_{135}\allowbreak\mathrm{CCZ}_{246}$ & $3$ & 48 & 12 & -- \\
6 & $\mathrm{CS}_{12}\allowbreak\mathrm{CS}_{34}\allowbreak\mathrm{CCZ}_{125}\allowbreak\mathrm{CCZ}_{136}\allowbreak\mathrm{CCZ}_{345}$ & $3$ & 48 & 12 & -- \\
6 & $\mathrm{CS}_{12}\allowbreak\mathrm{CCZ}_{123}\allowbreak\mathrm{CCZ}_{145}\allowbreak\mathrm{CCZ}_{246}$ & $3$ & 48 & 12 & -- \\
6 & $\mathrm{T}\otimes\allowbreak \bigl(\mathrm{CCZ}_{145}\allowbreak\mathrm{CCZ}_{235}\bigr)$ & $3$ & 47 & 12 & -- \\
6 & $\mathrm{T}^{\otimes 3}\otimes\allowbreak \mathrm{CCZ}$ & $3$ & 51 & 13 & -- \\
6 & $\mathrm{T}^{\otimes 6}$ & $3$ & 52 & 13 & -- \\
\addlinespace[3pt]
7 & $\mathrm{CS}_{12}\allowbreak\mathrm{CS}_{34}\allowbreak\mathrm{CCZ}_{125}\allowbreak\mathrm{CCZ}_{136}\allowbreak\mathrm{CCZ}_{247}\allowbreak\mathrm{CCZ}_{345}$ & $3$ & 48 & 13 & -- \\
7 & $\mathrm{CS}_{12}\allowbreak\mathrm{CCZ}_{123}\allowbreak\mathrm{CCZ}_{145}\allowbreak\mathrm{CCZ}_{146}\allowbreak\mathrm{CCZ}_{157}\allowbreak\mathrm{CCZ}_{245}\allowbreak\mathrm{CCZ}_{267}$ & $3$ & 48 & 13 & -- \\
\end{longtable}
\endgroup

\clearpage

\end{document}